\documentclass[longbibliography, amsmath, twocolumn, amssymb, aps, superscriptaddress]{revtex4-1}
\usepackage{xr}
\usepackage{amsmath}
\makeatletter
\newcommand*{\addFileDependency}[1]{
  \typeout{(#1)}
  \@addtofilelist{#1}
  \IfFileExists{#1}{}{\typeout{No file #1.}}
}
\makeatother
\usepackage{lineno}

\newcommand*{\myexternaldocument}[1]{%
    \externaldocument{#1}%
    \addFileDependency{#1.tex}%
    \addFileDependency{#1.aux}%
}
\myexternaldocument{Supplemental}

\usepackage{graphicx}   
\usepackage{dcolumn}  
\usepackage{bm}          
\usepackage{xcolor}    

\usepackage{hyperref}
\hypersetup{
    colorlinks=true,
    linkcolor=blue,
    filecolor=black,      
    urlcolor=blue,
    citecolor=blue,
}
\usepackage{natbib}
\usepackage{braket}
\usepackage{soul}
\usepackage{relsize}
            
\usepackage{pifont}

\begin{document}
\title{Emergence and Growth Dynamics of Wetting-induced Phase Separation on Soft Solids}

\author{Wenjie Qian}
\affiliation{Department of Physics, The Hong Kong University of Science and Technology, Hong Kong SAR, China}

\author{Weiwei Zhao}
\affiliation{Department of Physics, The Hong Kong University of Science and Technology, Hong Kong SAR, China}

\author{Tiezheng Qian}
\affiliation{Department of Mathematics, The Hong Kong University of Science and Technology, Hong Kong SAR, China}

\author{Qin Xu}
\email{qinxu@ust.hk}
\affiliation{Department of Physics, The Hong Kong University of Science and Technology, Hong Kong SAR, China}

\begin{abstract}

Liquid droplets on soft solids, such as soft polymeric gels, can induce substantial surface deformations, leading to the formation of wetting ridges at contact points. While these contact ridges have been shown to govern the rich surface mechanics on compliant substrates, the inherently divergent characteristics of contact points and the multiphase nature of soft reticulated gels pose great challenges for continuum mechanical theories in modeling soft wetting phenomena. In this study, we report {\em in-situ} experimental characterizations of the emergence and growth dynamics of the wetting-induced phase separation. The measurements  demonstrate how the migration of free chains prevents the stress singularities at contact points. Based on the Onsager variational principle, we present a phenomenological model that effectively captures the extraction process of free chains, including a crossover from a short-term diffusive state to a long-term equilibrium state. By comparing model predictions with experimental results for varied crosslinking densities, we reveal how the intrinsic material parameters of soft gels dictate phase separation dynamics.

\end{abstract}

\date{\today}
\maketitle


Wetting on soft solids plays essential roles in many emerging fields, including soft robotics~\cite{Shah2021}, flexible electronics~\cite{Sun2023}, cell patterning~\cite{Cavallini2012}, and bio-adhesive applications~\cite{Kim2021}. Compliant interfaces can deform significantly under the deposition of a liquid droplet, resulting in a ridge profile at contact lines~\cite{Didier1996}. Recent theoretical modelling~\cite{lubbers2014,Style2012_sm} and experimental characterizations~\cite{Park2014, Style2013, Zhao2022} have revealed the roles of contact ridges  beyond classical wetting theories. The static contact angles on soft solids vary with droplet sizes due to the rotation of wetting profiles, which differs from the prediction of Young--Dupre's law~\cite{Style2013, Style2013_PNAS}. Moreover, the dissipation originating from a moving contact ridge significantly slows down the spreading of liquid droplets~\cite{Karpitschka2015, Zhao2018, Xu2020}. Thus, the multi-scale responses of a contact ridge to droplet wetting is crucial for the rich mechanics at soft interfaces.

However, the  mechanical mechanisms underlying these wetting ridges remain unclear. Based on linear elastic theory, the bulk stress should diverge logarithmically with respect to the distance to contact points~\cite{Pandey2020, Limbeek2021}. While nonlinear models have been proposed to avoid the contact singularities~\cite{Robin2019, Dervaux2020}, their sufficiency for describing soft wetting profiles remains debated~\cite{Bain2021,heyden2023}. Beyond the framework of continuum mechanics, recent studies have focused on the multi-phase nature of soft polymeric gels~\cite{Xu2020, Jensen2015}. The wetting of liquid droplets on soft gels can potentially extract free chains from crosslinked networks and lubricate the solid--liquid interfaces~\cite{Aurelie2017,Aurelie2018}, providing an alternative mechanism to override the stress divergence~\cite{Jensen2015,Flapper2023}.  Although this wetting-induced phase separation has been observed on the soft substrates swollen by short polymers~\cite{Cai2021,Lukas2023}, it remains  challenging to experimentally characterize the  migrating dynamics of free chains. In addition, this extraction process is governed by dissipative non-equilibrium dynamics~\cite{Jinhwa2010}, which is also difficult to model theoretically.

In this work, we conducted {\em in-situ} confocal imaging of the migrating free chains induced by soft wetting,  and demonstrated how this process effectively prevents the stress divergence at contact points. Comparing the experimental results with a phenomenological model based on Onsager's variational principle~\cite{Doi_2011}, we uncovered the material parameters governing the growth dynamics of phase-separation regions. 
 
\begin{figure*}[t]
\centering
  \includegraphics[width=160mm]{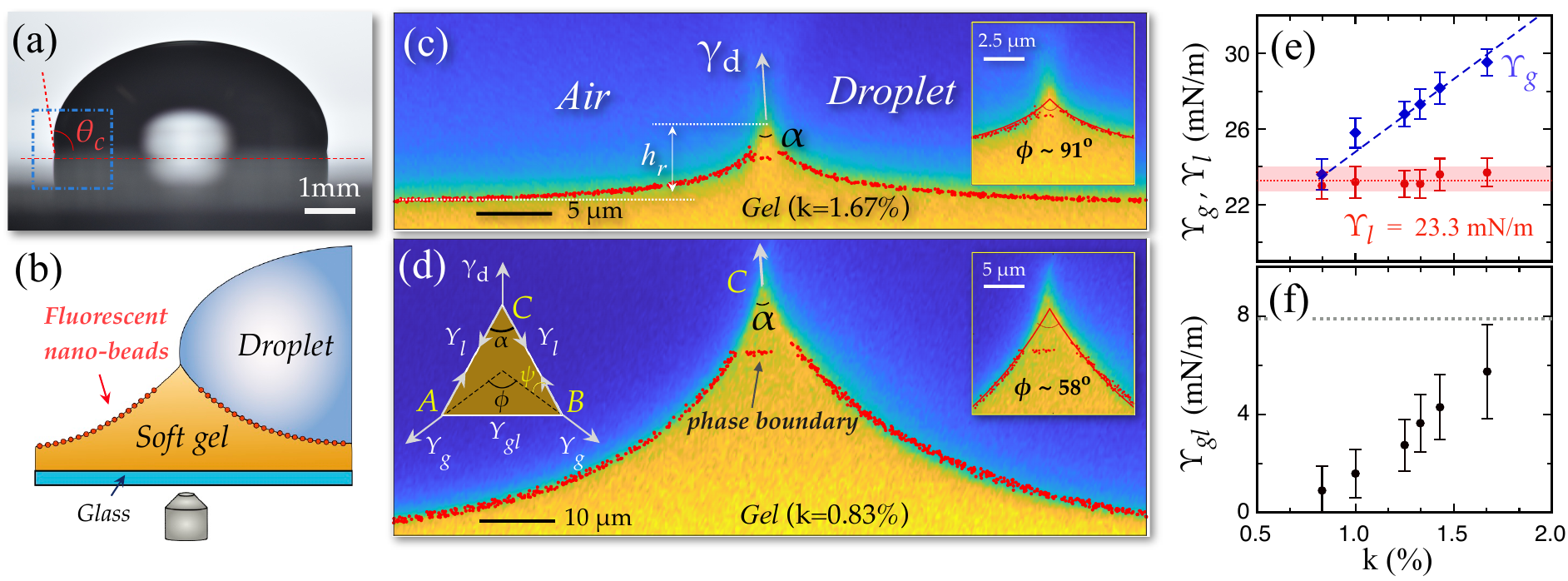}
  \caption{{\bf Local wetting profiles.} (a)~Snapshot of a glycerol droplet resting on a soft silicone gel with $k=1.43$~wt\%. The horizontal dashed line indicates the substrate surface. (b)~Schematic illustration of the imaging setup in our experiments. (c) \& (d)~Representative confocal images of phase separation induced by wetting on soft gels with $k = 1.67\%$ and $k = 0.83\%$, respectively. The inset images in panels (c) and (d) are expansions of areas in the images near contact points, where the red solid lines are the elastocapillary model with the best fitted $\Upsilon_g$.  The sketch in (d) indicates the local balance of surface stresses. (e)~Plots of $\Upsilon_g$ and $\Upsilon_l$ against $k$. The shaded red area indicates the measured $\Upsilon_l$ with uncertainties. (f)~Plot of $\Upsilon_{gl}$ against $k$. The gray dotted line is the upper limit of $\Upsilon_{gl}$ predicted by \cite{Jensen2015}. The error bars in both (e) and (f) represent the experimental uncertainties caused by determining the opening angles ($\alpha$, $\phi$ and $\psi$). }
  \label{fig:imaging}
\end{figure*}

{\it Materials and method.---} The soft  gels were prepared by mixing base polydimethylsiloxane (PDMS) polymers (Gelest, DMS-V31) with copolymer crosslinkers. The mixtures were coated onto glass cover-slips to form films with a thickness of approximately $45~\mu$m.  By adjusting the weight density of crosslinkers from $k=0.83\%$ to $2.0\%$, the shear modulus of soft gels varied from $G= 0.45$~kPa to $12.9$~kPa. Figure~\ref{fig:imaging}a shows a representative image of a glycerol droplet with a radius of $R_d=2.5$~mm, wetting on a gel substrate cured with $k=1.43\%$. The overall droplet shape  equilibrated within a minute after the deposition. By detecting the droplet boundary~\cite{Zhao2022}, we measured the apparent contact angle $\theta_c = 99.3^\circ \pm 0.6^\circ$, and the droplet surface tension $\gamma_d = 42.5\pm 0.8$~mN/m, which was markedly lower than the standard surface tension of liquid glycerol ($\sim 67$~mN/m). This disparity was an experimental indicator of the extracted free chains from soft gels covering the droplet~\cite{Aurelie2017, Aurelie2018, Zhao2022}, a phenomenon similar to the droplet cloaking process on lubricant-infused surfaces~\cite{Gunjan2021,Rodrique2022}.

To enable {\em in-situ} visualization of local wetting profiles, the bulk of soft gels were stained with a 0.1~wt\% UV-fluorescent dye (Tracer Product, TP3400-1P6). These oleophilic molecules were absorbed by the base PDMS polymers~\cite{Coux2020}. Simultaneously, a layer of 100~nm fluorescent nanobeads (FluoSpheres, Thermo Fisher) were deposited on gel surfaces. As these nanobeads preferentially interact with the crosslinked networks~\cite{Jensen2015}, they were utilized as the tracers to measure the network deformations. The UV  dye and the nanobeads were activated by 488~nm and 555~nm laser beams, respectively. We reconstructed the wetting profiles by combining the fluorescent signals acquired through different channels. 

{\it Local wetting profiles.---} Figures~\ref{fig:imaging}(c) and  (d) display the images of contact ridges for two crosslinking densities, $k=1.67\%$ and $k=0.83\%$, respectively. The red dots denote the locations of nanobeads and the yellow regions represent the gel matrix. The imaging was conducted three hours after the initial drop deposition to ensure equilibrium profiles. Due to the substantial difference between the shear moduli of the two gels ($9.5$~kPa and 0.45~kPa), the  ridge height on the softer substrate, $h_r = 26$~$\mu$m, was significantly higher than that on the stiffer substrate, $h_r = 4.2$~$\mu$m. Near the contact points, we observed a micro-sized triangular region labeled by the UV dye and bounded below by a line of fluorescent nanobeads. This feature resembles the geometries observed in phase separation induced by soft adhesion~\cite{Jensen2015} and fluid extractions on substantially swollen gels~\cite{Cai2021}. Since the emergence of this region was associated with the reduced droplet surface tension $\gamma_d =42.5$~mN/m, we concluded that the free chains had migrated to the contact point, where a phase  boundary between the gel and extracted chains was created. 

\begin{figure}[t]
\centering
	\includegraphics[width = 88 mm]{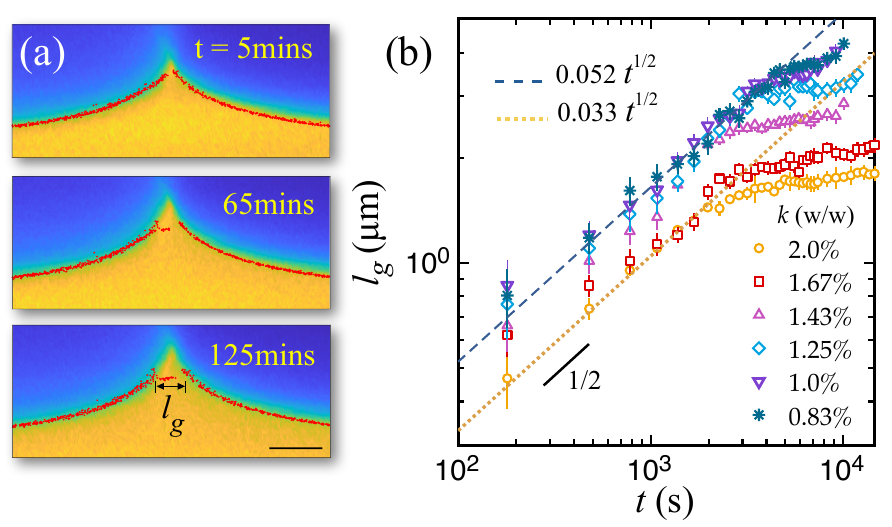}
	\caption{{\bf Growth of phase separation regions.} (a)~Snapshots of wetting profiles for $k=1.43~\%$ and imaged at three time points ($t=5$~mins,  65~mins, and 125~mins). Scale bar: 5~$\mu$m. (b)~Plots of $l_g$ against $t$ as the crosslinking density of soft gels increases from $k=0.83~\%$ to $k=2.0~\%$. The blue and orange dotted lines indicate the best fits to $l_g\sim t^{1/2}$ for $k=0.83\%$ and $k=2.0\%$, respectively.}
	\label{fig:dynamics}
\end{figure}

We denote surface stress as the force per unit length required to expand an interfacial area on soft gels, which may differ from their surface energy~\cite{Zhao2022, Xu2017}. The shape of the phase-separation regions determines the local balances of surface stresses~\cite{Jensen2015, Stefan2023}. As illustrated by the sketch in Fig.~\ref{fig:dynamics}(d), we designate $C$ as the contact point and $AB$ as the phase boundary.  The region of free chains ($ABC$) remained symmetric with respect to the droplet interface as the crosslinking density varied from $k=0.83~\%$ to $2.0~\%$, suggesting that the surface stress at the polymer-air interface was approximately equal to that at polymer-glycerol interface~\cite{Xu2017, Jacco_PRL_2018}. Thus, we simply used $\Upsilon_l$ and $\Upsilon_g$ to represent the surface stresses of free chains and gels, respectively,  and $\Upsilon_{gl}$ to denote the surface stress at the phase boundary ($AB$). 

The surface stress of free chains $\Upsilon_l$ and the droplet surface tension $\gamma_d$ form a classical Neumann's triangle at the contact point $C$~\cite{Style2013, Xu2017, Stefan2023}, where the opening angle of wetting ridges was found to be a constant, $\alpha = 48^\circ \pm 2^\circ$, independent of $k$. Considering $\gamma_d = 2\Upsilon _l  \cos(\alpha/2)$, we obtained a constant surface stress of free chains, $\Upsilon_l = 23.3\pm 0.4$~mN/m, which is close to the surface tension of pure liquid PDMS~\cite{Zhao2022}. The free chains effectively avoid stress singularities at contact points due to their negligible elasticity. Further, we determined the surface stress of soft gels $\Upsilon_g$ for various $k$ by fitting the profiles of nanobeads (red dots) to a linear elastocapillary theory (see Appendix B). For example, the solid red lines in the zoomed images in Figs.~\ref{fig:imaging}(c) and (d) are the best fits using $\Upsilon_g= 29.5$~mN/m and 23.6~mN/m, respectively, for the two different crosslinking densities.

Figure~\ref{fig:imaging}(e) summarizes how the measured surface stresses of soft gels ($\Upsilon_g$) and extracted polymers ($\Upsilon_l)$ vary with $k$. At the lowest crosslinking density $k =0.83~\%$, we observed that $\Upsilon_g \approx \Upsilon_l$, suggesting a negligible difference between the surface stress of the extracted PDMS chains and weakly crosslinked PDMS gels. For $k\geq 1~\%$, however, $\Upsilon_g$  becomes markedly higher than $\Upsilon_l$~\cite{Zhao2022}. Using the Neumann triangles at the endpoints $A$ and $B$, we estimated that $\Upsilon_{gl}=\Upsilon_g \sin(\phi/2)-\Upsilon_l\sin(\phi/2-\psi)$, where $\phi$ and $\psi$ are the relevant contact angles defined in the sketch in Fig.~\ref{fig:imaging}(d).  The resulting $\Upsilon_{gl}$ for different crosslinking densities is shown in Fig.~\ref{fig:imaging}(f). As $k$ varies from $0.83~\%$ to $2.0\%$, $\Upsilon_{gl}$ increases from $0.9$~mN/m to $5.7$~mN/m. These values remain below the estimated upper-bound of $\Upsilon_{gl}$ (gray dotted line), which was previously obtained from soft adhesion experiments~\cite{Jensen2015}.

{\it Growth dynamics of phase separation.---} The extraction of free chains was characterized by a slow dissipative process. For instance, Fig.~\ref{fig:dynamics}(a) exhibits representative snapshots of the wetting profile on a gel substrate with $k=1.43\%$, evolving over a period of about two hours. Since the region of extracted free chains grew in a self-similar manner (see  Fig.~\ref{Fig_geometry} in Appendix C), the lateral length of phase boundary ($l_g$) was measured tocharacterize this growth dynamics. Figure~\ref{fig:dynamics}(b) shows the plots of $l_g$ against $t$ as the crosslinking density increases from $k=0.83\%$ to $2.0\%$, where $t=0$ is defined as the moment at which the droplets were deposited. All traces of $l_g(t)$ follow a qualitatively similar trend: they first increase as $l_g\sim t^{1/2}$ in a short timescale and then ultimately plateau at an equilibrium length $l_{eq}$.  The crossover between the two regimes is determined by a characteristic timescale, $\tau_c\sim 10^{3}s$. For $k\leq 1.25~\%$, $l_g$ may continue to increase after reaching $l_{eq}$ for a certain period due to plastic deformations (see Fig.~\ref{Fig_plasticity} in the Appendix D). However, this work only focuses on the growth dynamics in linear elastic regimes.  Importantly, we found $l_{eq}$ to be independent of both droplet radius and substrate thickness~\cite{Suppl}, suggesting that the migration of free chains was governed by the material properties near contact points.

\begin{figure*}[t]
\centering
	\includegraphics[width = 160 mm] {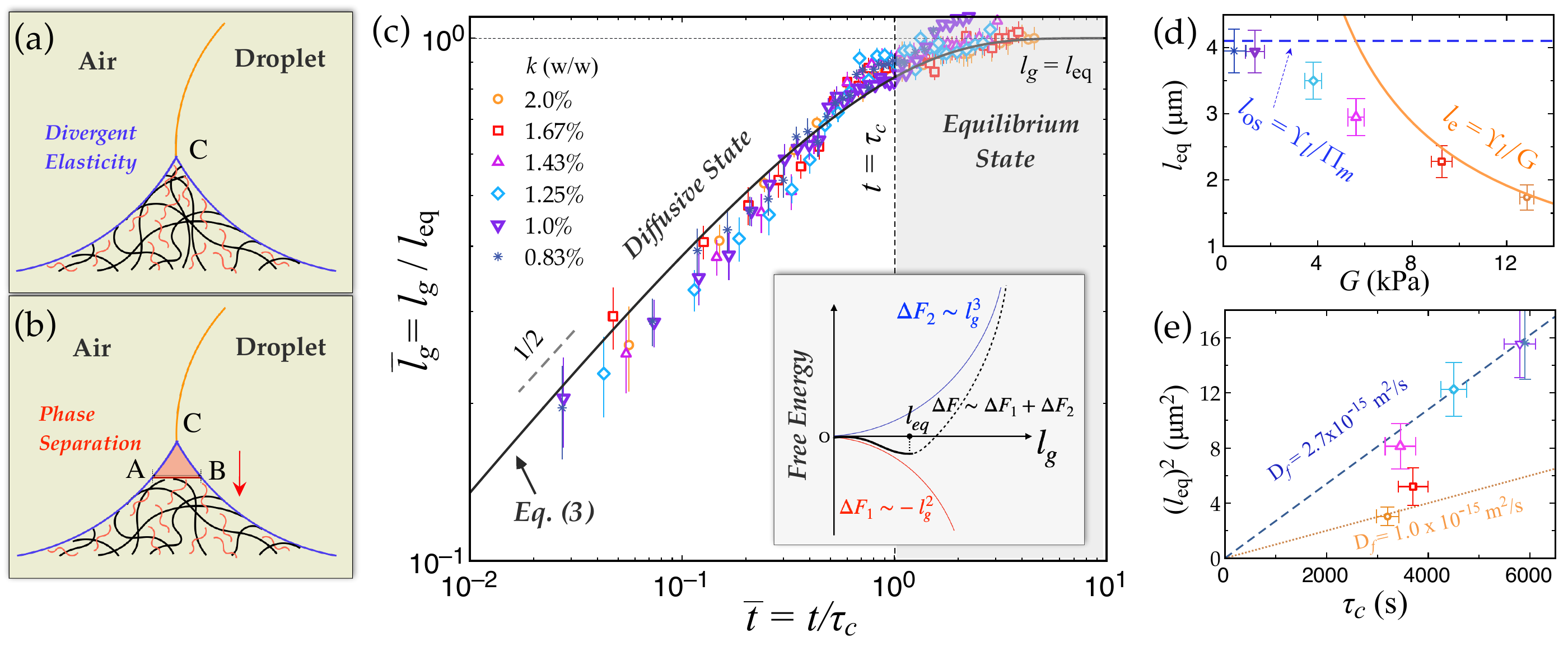}
	\caption{{\bf Growth dynamics of phase separation regions.} (a) Schematic of a homogenous wetting ridge with diverging elastic energy. (b) Schematic of a growing phase separation region. (c)~Collapse of the experimental results of $l_g(t)$ for different $k$ onto the prediction from Eq.~\ref{eq:master_curve}. Inset: the  free energy landscape $\Delta F$ results from the competition between $\Delta F_1$ and $\Delta F_2$. (d)~Plot of $l_{\rm eq}$ against the shear moduli ($G$) of soft gels with different $k$. The  dashed blue indicates the estimated osmocapillary length $l_{\rm os} = \Upsilon_l/\Pi_m$, while the solid orange line indicates the predicted elastocapillary length $l_{\rm e}=\Upsilon_l/G$. (e)~Plot of $l^2_{\rm eq}$ against $\tau_c$ obtained for different $k$. The blue dashed and orange dotted lines are replotted from the two $t^{1/2}$ scalings shown in Fig.~\ref{fig:dynamics}(b), which give rise to $D_f=2.7\times 10^{-15}$~m$^2$/s and  $D_f=1.0\times 10^{-15}$~m$^2$/s, respectively.}
\label{fig:analysis}
\end{figure*}

Here, we propose a phenomenological model to quantitatively describe the temporal evolution of $l_g$. We consider the free chains as the solvent and the crosslinked networks as the solute, which are mixed homogeneously in an undeformed  gel. When a glycerol droplet is deposited, the gel is first deformed instantaneously as a nearly incompressible solid, while the free chains have no time to migrate (Fig.~\ref{fig:analysis}(a)). The resulting wetting ridge creates a diverging elastic energy at the contact point, causing the system to become out-of-equilibrium. Consequently,  the crosslinked networks  retract downwards to reduce the elastic energy, but at an energetic cost of the phase-separation: the emergence of a solute-rich phase (below $AB$) and a solute-absent phase (above $AB$) (Fig.~\ref{fig:analysis}(b)). Due to the azimuthal symmetry of wetting profiles, this process can be modeled in a 2D plane. The area of the solute-absent phase is approximated as $A_f=\kappa l^2_g/2$, where $\kappa$ is a geometric constant. As the phase separation proceeds, the phase boundary $AB$ moves vertically downward at a velocity $v_I$. Given the mass conservation of free chains, $v_I l_g=dA_f/dt=d(\kappa l_g^2/2)/{dt}$, we obtain that $v_I=\kappa \dot{l}_g$.

We consider two factors that contribute to the change in free energy. First, assuming that $\beta$ is the energy density difference between crosslinked gels and free chains, the decrease in the free energy above $AB$ can be expressed as $ \Delta F_1 = - A_f \beta = -  \beta \kappa  l_g^2/2$. Since $\beta$ is supposedly dominated by the elasticity of networks near the contact point, $\Delta F_1$ represents the reduction in the elastic energy relative to the state of an initially deformed substrate (see Fig.~\ref{fig:analysis}(a)) in which the migration of free chains has not started yet ($l_g=0$). Second, the emergence of phase separation, driven by the decreasing elastic energy ($\Delta F_1$), creates an osmotic pressure ($\Pi$) that compresses the phase boundary $AB$. In a dissipative process, we assumed that $\Pi$ was consistently balanced by elastic stresses generated by the gel~\cite{Yamaue2004}, $\Pi \sim G {l_g}/{l_c}$, where $l_c$ is a phenomenological parameter to characterize the compressive strain. The increase in free energy of the crosslinked gels below $AB$ can be estimated as $\Delta F_2 = -\int \Pi dV \sim \int_0^{l_g} G (l_g^\prime/l_c) l_g^\prime d l_g^\prime = G l_g^3/(3l_c)$. Note that  $\Delta F_1$ and $\Delta F_2$ are both of elastic relevance but realized in different regions (above $AB$ and below $AB$) through different physical mechanisms, which eventually balance each other at equilibrium.

Thus, the change of total free energy becomes
\begin{equation}
\Delta F = \Delta F_1+\Delta F_2 \sim -\frac{1}{2} \beta\kappa l_g^2  +  \frac{1}{3} \frac{G}{l_c} {l_g^3}.
\label{eq:F}
\end{equation}
We neglected a linear term $\Delta F_s\sim\Delta\Upsilon l_g$, which represents the interfacial energy change due to the surface stress difference $\Delta\Upsilon=\vert \Upsilon_{gl}+\Upsilon_l-\Upsilon_{g} \vert$. The interfacial term $\Delta F_s$ can be important at small scales when $l_g < \Delta \Upsilon/G$. However, the relevant length scale $\Delta\Upsilon/G$ was significantly smaller than the experimentally measured $l_g$ for all different $k$, such that the impact of $\Delta F_s$ is negligible in our measurements. The free energy landscape of $\Delta F$ is schematically illustrated by the inset of Fig.~\ref{fig:analysis}(c).  Using $\partial \Delta F(l_g)/\partial l_g =0$, we obtain a finite equilibrium length, $l_{eq}=l_c \beta\kappa/G$, suggesting that the elasticity of crosslinked networks prevents an infinite growth of the phase separation region.

In a given cross-sectional area ($\Omega$), the dissipation between the networks and free chains with a relative velocity $v_p$ is expressed as $\Phi = (1/2)\int_\Omega \xi v^2_p~dxdy$, where $\xi$ is a frictional coefficient. Since the extraction of free chains is localized near contact points, we assume that $v_p = v_I g(x/l_g, y/l_g)=v_I g(\bar{x}, \bar{y})$, where $g(\bar{x}, \bar{y})$ vanishes asymptotically as $\bar{x},\bar{y} \to \infty$. Therefore, 
\begin{equation}
\Phi = \frac{1}{2} \int_\Omega \xi v^2_p dxdy = \frac{1}{2} \xi  \Delta \kappa^2 l_g^2 \dot{l}_g^2=\frac{1}{2} \zeta \kappa^2  l_g^2 \dot{l_g}^2
\label{eq:phi_2}
\end{equation}
where $\Delta=\int_\Omega g(\bar{x}, \bar{y})^2d\bar{x}d\bar{y} $ is a convergent integral and $\zeta = \xi \Delta$ indicates a global frictional constant.

To determine ${l}_g(t)$, we apply the Onsager variational principle, which states that the evolution of a system with the lowest dissipation is given by the minimum of the following Rayleighian function~\cite{Doi_2011}, $R(l_g,\dot{l}_g)  = \Phi + \Delta\dot{ F} = \frac{\beta\kappa}{l_{eq}} l_g^2 \dot{l}_g - \beta\kappa{l_g} \dot{l}_g +\frac{1}{2} \zeta \kappa^2 l_g^2 \dot{l}_g^2$. The derivative $\partial R/ \partial \dot{l}_g = 0$ yields a master curve for the growth dynamics:
\begin{equation}
\bar{t}+\bar{l}_g+\ln(1-\bar{l}_g)=0,
\label{eq:master_curve}
\end{equation}
where $\bar{l}_g = l_g/l_{eq}$ and $\bar{t} = \beta t/ (l_{eq}^2\zeta\kappa)$ are the reduced dimensionless parameters. For $\bar{t} \gg 1$, $\bar{l}_g = 1$, indicating the {\em equilibrium state} with $l_g=l_{eq}$. For $\bar{t}\ll1$, the leading order gives the scaling law $\bar{l}_g ={(2\bar{t})}^{1/2}$ for a {\em diffusive state}. At $\bar{t}=1$, the crossover timescale is given by $\tau_c = l^2_{eq}/(\beta/\zeta \kappa)$, or $\tau_c = l^2_{eq}/D_f$ with an effective diffusivity $D_f= \beta/(\zeta \kappa)$. Since the difference in energy density between crosslinked networks and free chains is supposedly determined by the gel elasticity, we conjecture that $\beta\sim G$. Given $\zeta$ as the flow resistance in soft gels, $D_f$ represents an effective poroelastic diffusivity~\cite{Xu2020, Flapper2023, Jinhwan2010,Joseph2020}.

With the best fitted $l_{eq}$ and $\tau_c$ for each $k$, we collapsed the measured $l_g(t)$ onto the prediction of Eq.~\ref{eq:master_curve} in Fig.~\ref{fig:analysis}(\textcolor{red}{c}) using the rescaled parameters $\overline{l}=l_g/l_{eq}$ against $\overline{t}=t/\tau_c$. Figure~\ref{fig:analysis}(d) shows the resulting $l_{eq}$ as a function of the shear moduli of different soft gels ($G$).  For  $G>4$~kPa, $l_{eq}$ decreases with $G$ and aligns with the estimation of the elastocapillary length, $l_{eq}\sim l_e = \Upsilon_l/G$. Since $l_{eq}=l_c \beta \kappa/G$,  the elastocapillary length also determines $l_c$, which indicates a characteristic length scale in compressed networks containing the flow of free chains~\cite{Berman2019}.

For $G\leq 1$~kPa, however, $l_{eq}$ remained consistently around  4~$\mu$m regardless of $G$. Here, we demonstrated that the {\em osmocapillary length} ($l_{os}$), as introduced by Liu and Suo~\cite{Liu2016},  accounts for this constant $l_{eq}$. By definition, $l_{os} = \Upsilon_l/\Pi_m$, where $\Pi_m$ represents the demixing-induced osmotic pressure. Based on the Flory--Huggin theory, we estimate $\Pi_m = - (k_B T/ v_c) (\ln\phi_f+1-\phi_f+\chi (1-\phi_f)^2)$ where $v_c$ and $\phi_f$ are the molecular volume and volume fraction of free  chains in soft gels, respectively, and $\chi$ is the Flory--Huggin parameter~\cite{Flapper2023, Flory1942}. For the PDMS investigated in this study, $v_c \approx 4.8 \times 10^{-26}$~m$^3$ and $\phi_f \approx 68~\%$ for $G \approx 1$~kPa (see Appendix A). As the free chains and crosslinked networks are chemically identical, we conjecture that  their affinity to free volumes are similar~\cite{Graessley2009}. By neglecting the term $\chi\phi_f^2$, we obtained that $\Pi_m \approx 6 $~kPa and $l_{os} \approx 4.1$~$\mu$m, as indicated by the blue dashed line in Fig.~\ref{fig:analysis}(d). The good agreement between $l_{os}$ and $l_{eq}$ for $G\leq 1$~kPa indicates that, in these ultrasoft polymeric gels, the demixing-induced osmotic pressure dominates over bulk elasticity within wetting ridges.  

Figure~\ref{fig:analysis}(e) shows the plot of $l_{eq}^2$ against $\tau_c$, where the slope indicates the poroelastic diffusivity $D_f = l_{eq}^2/\tau_c$. As $k$ increases from 0.83~$\%$ to 2.0~$\%$, $D_f$ is found to range between $1.0\times 10^{-15}$~m$^2$/s and $2.7\times 10^{-15}$~m$^2$/s. These measured values of $D_f$ are substantially lower than the previous estimations ($\sim 10^{-11}$ to $10^{-12}$~m$^2$/s) based on droplets lubricating or dewetting on different silicone gels (Dow Sylgard and   CY52-276)  with comparable viscoelasticity~\cite{Xu2020, Aurelie2018, Zhao2018_sm}. Additionally, the thermo-equilibrium of polymeric networks gives $D_f \sim G^{1/3} (k_BT)^{2/3}/\eta_0 \sim 10^{-13}$~m$^2$/s~\cite{Hu2011}, where the viscosity of base polymers  $\eta_0 = 0.98$~cSt. These discrepancies suggest that the bulk rheology of soft gels is inadequately predictive of phase separation dynamics.

To determine the origin of this small diffusivity ($D_f \sim 10^{-15}$~m$^2$/s), we subjected soft PDMS gels to a two-week swelling process using ethanol--toluene mixtures. The uncrosllinked free chains were extracted by the swelling mixtures, and then collected for further rheological characterizations after drying the volatile components in solvents. Unlike the Newtonian base polymers ($\eta_0 \approx 1$~Pa$\cdot$s), the extracted polymers exhibited a shear thinning behavior with a zero-shear viscosity of approximately $\sim 10^3$~Pa$\cdot$s for varying crosslinking densities, which is roughly three orders of magnitude higher than $\eta_0$ (see Fig.~\ref{Fig_freechains} in Appendix A). This result suggests that, unlike the base PDMS polymers, the extracted polymers were likely made of partially crosslinked chains unattached to gel networks. Thus, we attribute the small $D_f$ shown in Fig.~\ref{fig:analysis}(c) to the enhanced frictional resistance between those partially crosslinked chains and fully crosslinked networks.

In a conventional indentation setup equipped with a millimeter-scale probe ($L\sim 1$~mm)~\cite{Hu2010,Kalcioglu2012}, the compression-induced extraction of free chains will take an extraordinary long  period, $t\sim L^2/D_f \sim 10^9$~s. This explains why the specific silicone  gels (Gelest DMS-V31) investigated in this work exhibited solely viscoelastic responses to macroscopic indentations~\cite{Xu2020}. In soft wetting, however, the free chains could diffuse through micron-sized contact ridges within hours (Fig.~\ref{fig:dynamics}), allowing us to directly characterize the poroelastic diffusion.

{\it Conclusions.---} In summary, we conducted {\em in-situ} measurements of the emergence and slow growth dynamics of wetting-induced phase separation  on soft PDMS substrates (Fig.~\ref{fig:imaging} and Fig.~\ref{fig:dynamics}). We  confirmed that the extracted free chains preserve the local balance of surface stresses and avoid the divergence of bulk elasticity (Fig.~\ref{fig:imaging}). Based on Onsager's variational principle, we proposed a phenomenological model with minimal assumptions to describe the development of phase separation. The model presents an explicit ansatz (Eq.~\ref{eq:master_curve}) that captures the crossover from a short-term diffusive to a long-term equilibrium state. By comparing the experimental results with the theoretical predictions, we identified the key material parameters that govern the phase-separation process at different stages (Fig.~\ref{fig:analysis}). These findings will benefit the design of functional soft interfaces, such as lubricant-infused polymeric substrates, for controlling droplet wetting. 

We thank Prof.~Ni Ran for useful discussions. W.~Qian, W.~Zhao and Q.~Xu were supported by the General Research Funds (16305821 and 16306723), the Early Career Scheme (26309620), and the Collaborative Research Fund (C6004-22Y) from the Hong Kong Research Grants Council. T.~Qian was supported by the Collaborative Research Fund (C1006-20WF) and General Research Fund (16306121) from the Hong Kong Research Grants Council. We also appreciate the funding support from the GDST Collaborative Research Grant (2023A0505030017).

\appendix
\renewcommand{\appendixname}{APPENDIX}
\section{MATERIAL DETAILS}

The soft polymeric gels examined in this study were comprised of polydimethylsiloxane (PDMS) base polymers (Gelest DMS-V31) crosslinked by trimethylsiloxane terminated-copolymers (Gelest, HMS-301). The crosslinking processes were catalyzed by a platinum divinyltetramethyldisiloxane complex (Gelest, SIP6831.2). The molecular weight of the base PDMS polymers is $M_c \approx 2.8 \times 10^4$~g/mol, corresponding to a molecular volume $v_c \approx 4.8 \times 10^{-26}$~m$^3$.

\begin{figure}[h]
  \centering
    \includegraphics[width=60mm]{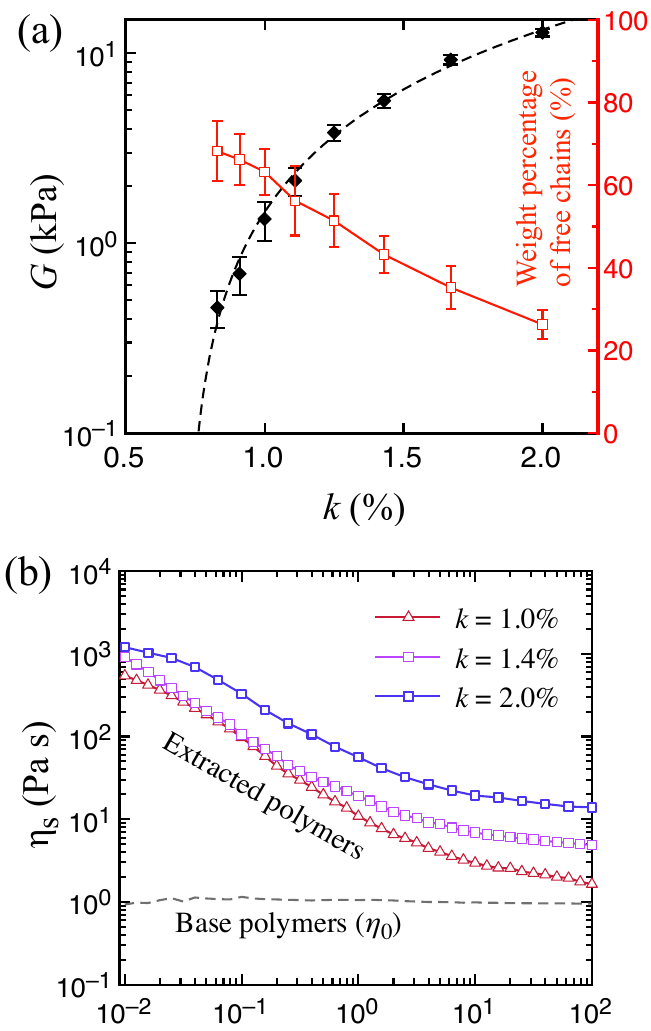}
    \caption{\textbf{Material properties of the soft silicone gels.} (a) Plots of the shear modulus of soft silicone gels $G$ (black solid diamonds) and the weight percentage of free chains (red hollow squares) against $k$. The black dashed line indicates the best fits, $G(k) = G_0 (k-k_0)^n$, with $G_0= 9.0$~kPa, $k_0=0.72~\%$, and $n=1.47$. (b) Flow curves of the extracted polymers from gels with different crosslinking densities ($k=1.0~\%$, 1.4~\%, and 2.0~\%). Additionally, the viscosity of the base polymers is  plotted as a reference.}
    \label{Fig_freechains}
\end{figure}

Prior to curing, we prepared a Part $A$ solution consisting of the polymer base mixed with $0.01~\%$~wt catalyst, and a Part B solution containing the polymer base mixed with $10~\%$~wt crosslinkers. To synthesize PDMS gels with varying shear moduli, we controlled the mixing ratio of Part $A$ and Part $B$, giving a crosslinking density $k$ defined as the weight percentage of the crosslinkers. We allowed 48 hours for complete curing of the gels after mixing all of the chemical components. The shear moduli of the resulting soft silicone gels ($G$) with different crosslinking densities ($k$) are represented by the black solid diamonds in Fig.~\ref{Fig_freechains}(a). The experimental data can be fitted by $G(k) = A (k-k_0)^n$ with $G_0= 9.0$~kPa, $k_0=0.72~\%$, and $n=1.47$. The critical crosslinking density $k_0$ represents the minimum amount of crosslinkers required to generate a reticulated gel network. For $k<k_0$, the mixtures result in viscoelastic fluids instead of percolated gels~\cite{Zhao2022}.

To characterize the rheology of free chains, we gradually swelled the gels using a toluene ($40\%$~w/w)-ethanol mixture ($60\%$~w/w). We prepared a millimeter-sized cylindrical PDMS gel, which was then immersed in the aforementioned mixture for a week. The swelling solvent was replaced with a new batch daily, and we cumulatively collected the used batch. After the volatile components had been completely evaporated, we obtained the free chains extracted from the soft gels. The weight percentages of extractable free chains for different $k$ are shown by the red hollow squares in Fig.~\ref{Fig_freechains}(a). To demonstrate the rheological properties of free chains, Figure~\ref{Fig_freechains}(b) exhibits the plots of $\eta{(\dot\gamma})$ for soft gels with different crosslinking densities ($k=1.0~\%$, 1.4~\%, and 2.0~\%), together with the flow curve of the base polymers ($\eta_0$). The shear thinning behaviors of extracted polymers were caused by shear-induced disentanglement. 

\begin{figure*}[ht]
  \centering
    \includegraphics[width=165mm]{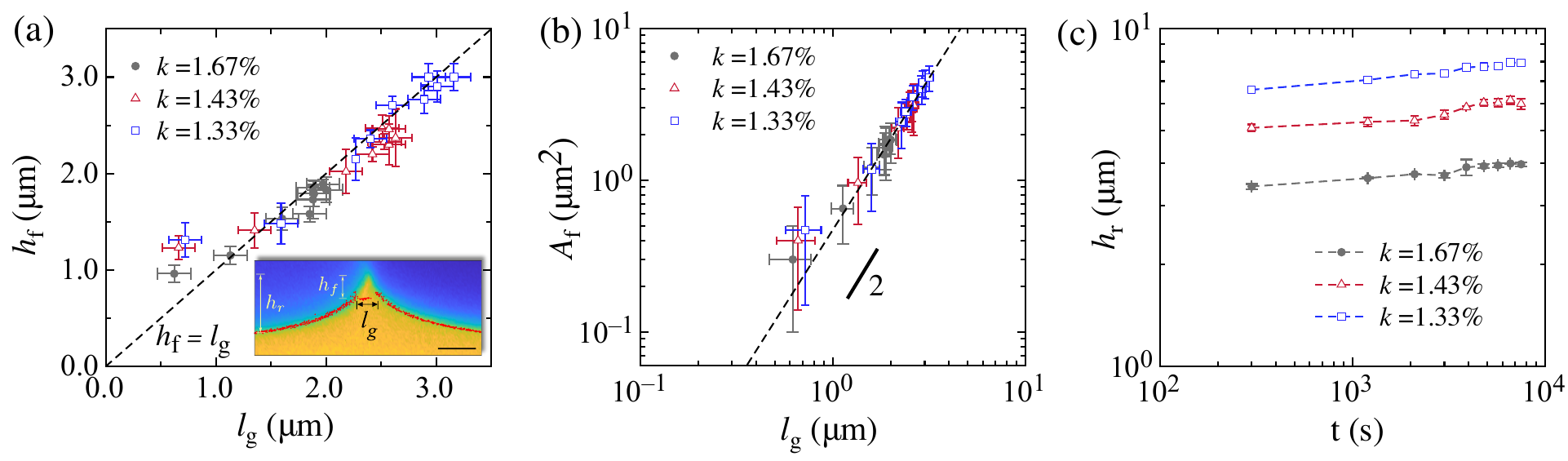}
    \caption{{\bf Geometry of the phase separation regions.}  (a) \& (b)  Plots of $h_f$ and $A_f$ measured at different time points  against $l_g$ for various crosslinking densities $k$.  The inset in (a) illustrates the definitions of the geometric parameters: $l_g$, $h_f$, and $h_r$. The dashed line in panel (a) indicates $h_f = l_g$, and the dashed line in panel (b) indicates the best fit of $A_f = \frac{1}{2} \kappa l_g^2$ with $\kappa=0.92 \pm 0.02$. (c) Plot of the total ridge height $h_r$ against time for different crosslinking densities.}
    \label{Fig_geometry}
\end{figure*}

\section{LINEAR ELASTOCAPILLARY MODEL}

To determine the surface stress of soft gels ($\Upsilon_g$), we fitted the experimental profiles of fluorescent nanobeads to a linear elastocapillary model~\cite{Zhao2022}. The governing equations of the displacement $({\bf u}(r,z))$ and stress tensor $({\boldsymbol\sigma}(r,z))$ of the soft substrate are expressed as
\begin{align}
& (1-2\nu) \nabla^2 {\bf u} + \nabla(\nabla\cdot {\bf u})=0,\\
&{\boldsymbol\sigma} = \frac{2}{1+\nu} [\frac{1}{2} ((\nabla {\bf u})^T + \nabla {\bf u})+\frac{\nu}{1-2\nu}(\nabla\cdot {\bf u}) {\bf I}],
\end{align}
with the boundary conditions
\begin{widetext}
\begin{align}
& \sigma_\Upsilon = \Upsilon_g \frac{1}{r} \frac{\partial}{\partial r}  (r\frac{\partial u_z}{\partial r})\hat{z}, \\
& t(r,z=h) =  \gamma_d \sin{\theta} \delta (r-R\sin\theta) \hat{z} - \frac{2\gamma_d}{R_d} H(R_d\sin\theta - r) \hat{z} - \gamma_d\cos\theta\delta(r-R_d\sin\theta) \hat{r}.
\end{align}
\end{widetext}
By applying the Hankel transformations to both the displacement $u_z(r,z)$ and stress field $\sigma(r,z)$, we obtain the surface profile
\begin{widetext}
\begin{equation}
    \begin{aligned}
u_z(r,h) = 
&\int_0^{+\infty} ds\   {\gamma_l} s J_0(sr) \\ 
&({J_1(sR\sin\theta)} s (\nu+1) \cos\theta \left(2
   h^2 s^2+(2 (5-4 \nu) \nu-3) \cosh (2 h s)+2 \nu (4 \nu-5)+3\right)\\ 
   &+2  {J_0(sR\sin\theta)} R s \left(\nu^2-1\right) \sin\theta ((4 \nu-3) \sinh
   (2 h s)+2 h s)\\ 
   &-4 {J_1(sR\sin\theta)} \left(\nu^2-1\right) ((4 \nu-3)
   \sinh (2 h s)+2 h s)) \\
    &/(s^2 (E \left(2 h^2 s^2+4 \nu
   (2 \nu-3)+5\right)+E (3-4 \nu) \cosh (2 h s)\\ 
   &+4 {\Upsilon_g} h s^2
   \left(\nu^2-1\right)+2 {\Upsilon_g} s (\nu-1) (\nu+1) (4 \nu-3) \sinh
   (2 h s))),
   \label{eq:integral}
    \end{aligned}
\end{equation}
\end{widetext}
where $J_0$ and $J_1$ are zeroth-order and first-order Bessel functions. The red solid lines in the insets of Figs.~\ref{fig:imaging}(c) and (d)  are the best fits of Eq.~\ref{eq:integral} to the experimental results.

\section{PHENOMENOLOGICAL MODEL}
\subsection {Change of free energy in phase separation}

We consider the emergence of free chains as a phase separation process. Based on the variational nature of Onsager’s principle, our phenomenological model is a reduced description using the lateral length of phase boundary ($l_g$) as the only state variable  whose time evolution is directly measured in our experiments. We consider the free chains as solvent and the crosslinked networks as solute. In equilibrium states, the two components are homogeneously mixed in an undeformed gel. However, when a wetted glycerol droplet is deposited on the soft gel, the resulting capillary ridges induce diverging elastic energy, causing the system to be out-of-equilibrium. To re-equilibrate,  the crosslinked networks retract to reduce elastic energy ($\Delta F_1$), resulting in a region of free chains near the contact point. Meanwhile, the decomposition process is  associated with the work done by overcoming the osmotic pressure at the phase boundary $AB$ ($\Delta F_2$). Since $\Delta F_1 \sim -l_g^2$ and $\Delta F_2 \sim l_g^3$  (Eq.~\ref{eq:F}), respectively,  the competition between the two factors gives rise to an equilibrium length of the phase separation region (as shown in the inset of Fig.~\ref{fig:analysis}(c)).

\subsection {Geometry of the phase-separation regions}

We used the term $\frac{1}{2} \kappa l_g^2$ to estimate the area of phase separation regions ($A_f$), where the geometric constant $\kappa$ indicates  how well these domains can be estimated as equilateral triangles. For instance, a perfect equilateral triangle would lead to $\kappa = 1$. In the experiments, we first measured the height of phase separation regions ($h_f$) versus $l_g$ for  $k=1.33\%$, $1.43\%$, and $1.67\%$. Figure~\ref{Fig_geometry}(a) indicates that $l_g$ and $h_f$ are approximately equal in all the measurements, suggesting that the phase separation grew in a self-similar fashion. Further, Fig.~\ref{Fig_geometry}(b) shows the  plot of $A_f$ against $l_g$ for different crosslinking densities,  where all the points are nicely fitted to $A_f = \frac{1}{2}\kappa l_g^2$ with $\kappa = 0.92 \pm 0.02$. 

Additionally, we characterized how the total ridge height $h_r$ evolved over time. Figure ~\ref{Fig_geometry}(c) shows the plots of $h_r(t)$ for $k=1.33\%$, $1.43\%$, and $1.67\%$, respectively. Over the two-hour measuring period, $h_r$ remains primarily unchanged for various crosslinking densities. This result suggests that  the viscoelastic relaxation of gel networks  played a negligible role during the growth of the phase separation regions~\cite{Karpitschka2015}.  

\begin{figure*}[t]
  \centering
    \includegraphics[width=170mm]{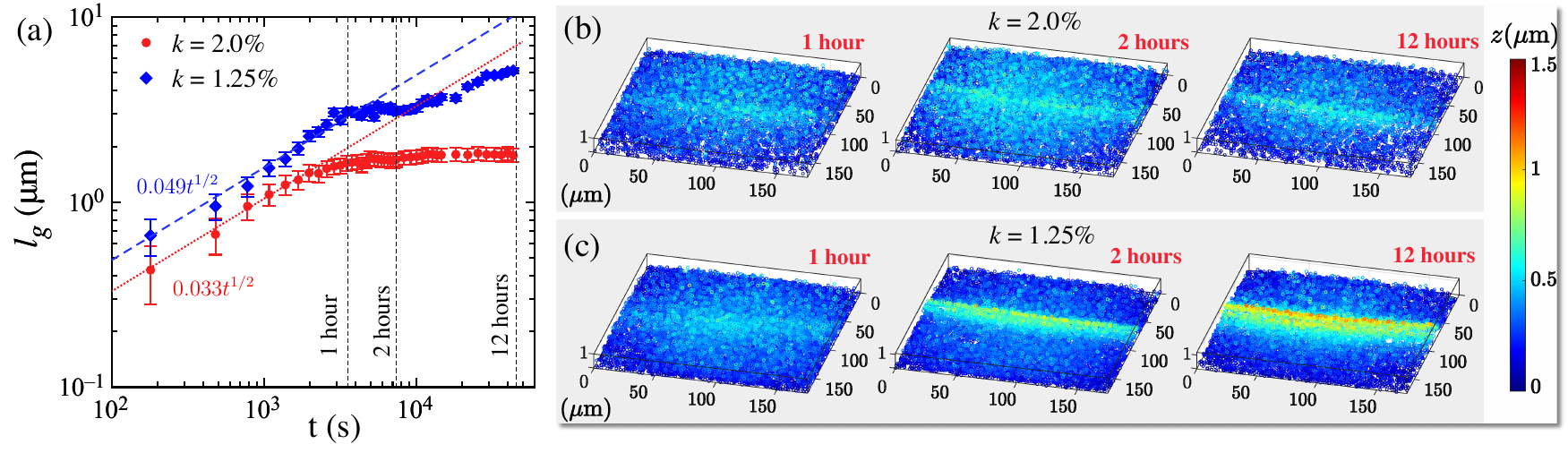}
    \caption{{\bf Wetting-induced plastic deformations on soft gels.} (a) Plots of $l_g$ against $t$ measured over a period of 12~hours for $k=1.25\%$ and $k=2.00\%$, respectively. (b) Surface profiles of soft gels with $k=2.00\%$ after removing the droplet-air interface at $t=1$~hour, $2$~hours, and 12~hours, respectively. (c) Surface profiles of soft gels with $k=1.25\%$ after removing the droplet-air interface at $t=1$~hour, $2$~hours, and 12~hours, respectively. }
    \label{Fig_plasticity}
\end{figure*}

\subsection {Viscous dissipation}

In the framework of  Onsager's variational principle, we need an explicit expression of dissipation function $\Phi$ to derive the growth dynamics of phase separation.  When the free chains migrate through the crosslinked networks, the density of dissipative energy is given by
\begin{equation}
\phi = \frac{1}{2} \xi v_p^2,
\label{dissipation_density}
\end{equation} 
where $\xi$ is a frictional coefficient and $v_p$ is the relative velocity between free chains and crosslinked network. As a result, the total dissipation can be calculated as 
\begin{equation}
\Phi = \int_{\Omega} \phi dx dy = \frac{1}{2} \int_{\Omega}\xi v_p^2 dx dy,
\label{dissipation}
\end{equation}
where $\Omega$ indicates the cross section area of wetting ridges. We assume that the migration of free chains is concentrated near the contact point, such that $v_p$ vanishes as $x, y \gg l_g$.  Therefore, the generic formula of  $v_p$ can be  written as
\begin{equation}
v_p = v_I g(\frac{x}{l_g}, \frac{y}{l_g}),
\label{v_p}
\end{equation}
where the dimensionless function $g(x/l_g, y/l_g)$ approaches zero as $x$ or $y \gg l_g$.  Note that $v_I = \kappa \dot{l}_g$ is the moving speed of the phase boundary $AB$. Combining Eq.~\ref{dissipation} and Eq.~\ref{v_p}, we obtain 
\begin{align}
\notag\Phi & =  \frac{1}{2} \xi v_I^2 \int_\Omega [g(x/l_g, y/l_g)]^2 dx dy \\
	& = \frac{1}{2} \xi \kappa^2  \dot{l}_g^2 l^2_g \int_{\bar{\Omega}} [g(\bar{x}, \bar{y})]^2 d\bar{x} d\bar{y} \;\;\:\; (\bar{x} = x/l_g, \bar{y} = y/l_g).
\end{align}
Since the total dissipation can not be divergent, we assume $\Delta = \int_{\bar{\Omega}} [g(\bar{x}, \bar{y})]^2 d\bar{x} d\bar{y}$ as a converging constant, such that
\begin{equation}
\Phi = \frac{1}{2} \xi \Delta \kappa^2  l_g^2 \dot{l}_g^2,
\end{equation}
which is consistent with the expression in Eq.~\ref{eq:phi_2}.

\subsection {Derivation of the mastering curve}

Here, we derive the mastering curve (Eq.~\ref{eq:master_curve}) governing the growth dynamics of the phase separation region. Given the Rayleighian function  $R(l_g,\dot{l}_g)  = \Phi + \Delta\dot{ F} = ( \frac{\beta\kappa}{l_{eq}} l_g^2 \dot{l}_g - \beta\kappa{l_g} \dot{l}_g +\frac{1}{2} \zeta \kappa^2 l_g^2 \dot{l}_g^2)$, the Onsager variational principle leads to 
\begin{equation}
\frac{\partial  R(l_g,\dot{l}_g)}{\partial \dot{l}_g}=\kappa l_g (\frac{\beta}{l_{eq}}l_g-\beta+\zeta \kappa l_g\dot{l}_g) = 0.
\label{eq:partial_l}
\end{equation}
By rearranging the Eq.~\ref{eq:partial_l}, we obtain 
\begin{equation}
\beta dt =\frac{\kappa\zeta dl_g}{1/l_g - 1/l_{eq}}.
\label{eq:diff}
\end{equation}
Integration of both sides of Eq.~\ref{eq:diff} gives 
\begin{equation}
\beta t=l_{eq}\zeta\kappa (-l_g+l_{eq}\ln\frac{l_{eq}}{l_{eq}-l_g}). 
\label{eq:analytic}
\end{equation}
By defining $\bar{t} = t\beta/(l^2_{eq}\zeta\kappa)$ and $\bar{l}_g=l_g/l_{eq}$, Eq.~\ref{eq:analytic} can be rewritten as $\bar{t}+\bar{l}_g+\ln(1-\bar{l}_g)=0$, consistent with the expression in Eq.~\ref{eq:master_curve}.

\section{WETTING-INDUCED SURFACE PLASTICITY}

We also conducted experimental characterizations to examine whether the presence of wetting ridges causes plastic deformations on gel surfaces. Figure~\ref{Fig_plasticity}(a) exhibits the temporal evolution of $l_g$ for $k=1.25\%$ (blue diamonds) and $k=2.00\%$ (red circles) over a period of 12 hours. For both crosslinking densities, we observed a diffusive scaling $l_g \sim t^{1/2}$ at the early stage, followed by a plateau value of $l_g$. For $k=2.00~\%$, $l_g$ remained primarily constant within the plateau regime. In contrast, for $k=1.25\%$, $l_g$ was found to  further increase  beyond $t=2$~hours after briefly reaching the plateau value. In our experiments, this second increasing stage of $l_g$ was commonly observed for soft gels with a low crosslinking density, $k \leq 1.25\%$.

We interpret the long-term increase of $l_g$ as a signature of irreversible plastic deformations. To confirm this assumption, we removed the droplet-air interface at different time points by adding excessive liquid glycerol  and characterized the fully relaxed surface profiles. Figure~\ref{Fig_plasticity}(b) shows the completely relaxed surface profiles of soft gels with $k=2.00\%$ after removing the droplet-air interface at $t=1$~hour, $2$~hours, and $12$~hours. The out-of-plane deformations remain mostly below 0.5~$\mu$m, close to the z-resolution of confocal imaging. In contrast, for a soft gel with $k = 1.25$~\%, the remaining surface deformations were significant ($>1$~$\mu$m) when the droplet-air interface was removed at $t=2$~hours or $12$~hours (Fig.~\ref{Fig_plasticity}(c)). These unrelaxed profiles represent irreversible plastic deformations at contact lines induced by droplet wetting. In this work, we only focused on the linear elastic regime and did not discuss the growth dynamic of phase separation in the plastic regime.

\bibliography{reference.bib}

\end{document}